\documentclass[
 aps,
 prl,
 twocolumn,
]{revtex4-1}

\usepackage{amsmath}
\usepackage{graphicx}
\usepackage[outdir=./figures/]{epstopdf}
\usepackage[usenames,dvipsnames,svgnames,table]{xcolor}
\usepackage[utf8]{inputenc}
\usepackage{hyperref}
\usepackage{pbox}
\usepackage{footmisc}
\usepackage[english]{babel}
\begin{document}

\title{Impact of Coulomb Scattering on Argon Plasma Based Thermionic Converter Performance}

\author{ R. E. Groenewald}
\affiliation{Modern Electron Inc. Bothell, WA 98011, USA}

\date{\today}

\begin{abstract}

Particle-in-cell (PIC) simulations were performed to study the impact of Coulomb scattering on the performance of argon plasma based thermionic converters. Using a simplified model, studies from the 1970's have concluded that plasma resistance, brought on by Coulomb collisions, causes a shift in the IV-curves of thermionic converters that use an argon plasma to mitigate space charge, thereby strongly limiting their electricity generation capability. In this work the impact of Coulomb collisions in such devices were studied as a function of the relative electrical potential between the electrodes, with higher fidelity through the use of a fully kinetic approach (PIC). This revealed that earlier reports overestimated the negative impact of Coulomb collisions around the flat-band potential. The results of the simulations are also used to comment on the validity of the assumptions made in the simplified model.

\end{abstract}

\maketitle

\subsection{Introduction}

A thermionic electrical converter (TEC) is a type of heat-to-electricity converter \cite{Schlichter,hatsopoulos_thermionic_1973,hatsopoulos_thermionic_1979}. Historically TECs were used for electricity generation in satellites using nuclear reactors since they are particularly well suited to high temperature ($>1300$ $^o$C) applications\cite{gyftopoulosThermionicNuclearReactors1963,benke1994operational,gryaznov200030th}. The lack of moving parts in these converters as well as their relatively high conversion efficiency (relative to other heat engines) at the extreme temperatures achievable with nuclear reactors makes them particularly attractive for space applications\cite{clarkSolarThermionicTest2006}. Recently, however, there has been renewed interest in these devices for terrestrial electricity generation\cite{fitzpatrick1997updated,goThermionicEnergyConversion2017,patent:20200294779}, which generally requires lower temperature operation ($< 1200$ $^o$C).
\begin{figure}[b]
	\includegraphics[width=0.85\columnwidth]{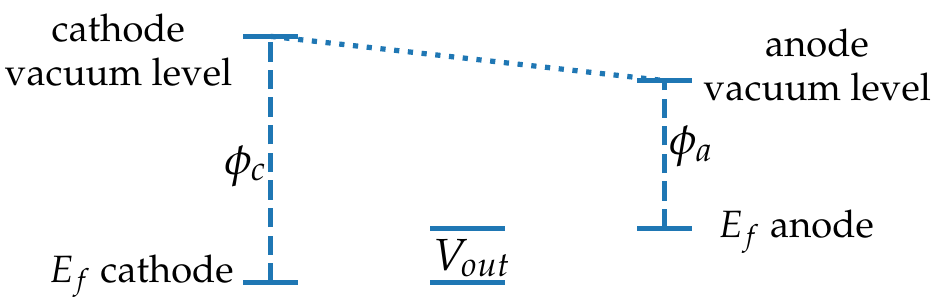}
	\caption{Schematic of the electron motive in a vacuum diode to show relation of different physical parameters of interest in a thermionic diode. $E_f$ indicates Fermi levels, $\phi_c$ is the cathode work function, $\phi_a$ the anode work function and $V_{out}$ the output voltage of the device.}
\label{fig:energy_diagram}
\end{figure}
The basic operation of a thermionic converter is as follows: The converter consists of a hot electrode (cathode) from which electrons are thermionically emitted\cite{https://doi.org/10.1002/andp.18822520602}. Separated from the cathode by some distance, referred to as the inter-electrode gap, is another electrode at a colder temperature (anode) on which the thermionically emitted electrons are collected. If these two electrodes are connected with an external conductor a current will flow through that conductor. The schematic in Fig.~\ref{fig:energy_diagram} shows the electron motive (defined as $-eV$ where $V$ is the electrical potential and $e$ the electronic charge) for a vacuum diode. From this diagram it can be seen that a thermionic diode can be used to perform electrical work (generate electricity) if $\phi_c - \phi_a - V_{out} > 0$, where $\phi_c$ is the cathode work-function, $\phi_a$ the anode work-function and $V_{out}$ the potential difference at which the two electrodes are held. A common issue with TECs is the so-called space charge problem which arises due to the Coulomb repulsion between electrons as they transit the inter-electrode gap, which limits the current flow through the device\cite{PhysRevSeriesI.32.492,PhysRev.2.450}. Frequently a cesium plasma has been used to neutralize the inter-electrode region and thereby mitigate the space charge problem\cite{1444197,rasor1991}. This works well for TECs operating at high temperature since the high temperature of the emitted electrons allow a sustained arc-discharge between the cathode and anode, the so-called ignited mode\cite{wilkinsThermionicConvertersOperating1966}. However, at operating temperatures below approximately $1200$ $^o$C this mode of operation limits the conversion efficiency achievable in a TEC due to inefficient plasma generation. Alternatively, inert gas plasmas have also been used in place of the Cs plasma. Due to the higher ionization energy for inert gas atoms compared to Cs atoms, a sustained arc-discharge is not achievable in the power producing mode. Therefore, a different strategy has to be used to generate the plasma. One way to do this is to apply short high voltage pulses across the diode during which the electrons in the inter-electrode gap are accelerated to sufficiently high velocity to impact ionize neutral atoms\cite{mcveyImprovedPulsedIonization1990,PhysRevE.103.023207}. Another is to add a third hot electrode (an auxiliary electrode)\cite{4317090,Limpaphayom} that also thermionically emits electrons and is held at a sufficiently negative bias relative to the cathode and anode so that the electrons emitted from this extra electrode can impact ionize neutral atoms as they transit through the inter-electrode gap. This article focuses on the latter approach. Hansen et. al. (1976)\cite{plasmatron} also studied this approach and generally refer to a TEC that relies on an auxiliary electron emitting electrode to generate a plasma in the inter-electrode gap as a "plasmatron". In Ref.~\cite{plasmatron} they describe experimental results they obtained with a specific plasmatron design that was comprised of a heated filament placed in the inter-electrode gap of a thermionic diode. They also discuss a theoretical model to describe the performance of a generic plasmatron device, which is discussed below. The generic device is assumed to have some rate of plasma formation in the inter-electrode gap that is independent of the output voltage of the thermionic diode. They concluded that plasma resistance is a limiting factor in the energy conversion potential of the plasmatron. It is this conclusion that will be explored and challenged in the present article. Importantly, Hansen et. al (1976) treated Coulomb scattering (plasma resistivity) as a pure voltage shift in the JV-curve (current density versus voltage bias) that characterizes the plasmatron, but in reality the system should be self-consistently solved, since the inclusion of Coulomb scattering alters the electron temperature and steady state plasma density. For this reason, PIC simulations, which can capture these effects, are used in the present study of the system as described in the Model section below.

\subsubsection{Simplified Plasmatron Model}

The simplified model developed by Hansen et. al. (1976)\cite{plasmatron} to study the plasmatron performance is described in detail in that reference and the interested reader is directed there for a study of the model. In their introduction of the model, Hansen et. al. noted that the typical operating regime of the plasmatron with a low pressure of argon (such that the electron and ion mean free paths are of the same order as the inter-electrode gap), makes it difficult to study the system with effective models. There are too many collisions to assume a collisionless (Knudsen) plasma but too few to justify the use of continuum transport equations - note that this regime is an ideal use case for PIC simulations\cite{Tskhakaya}. Nevertheless Hansen et. al. employ the continuum transport equations derived by Ecker (1964)\cite{Rosenbluth1964AdvancedPT} to study the plasmatron performance. Based on the recognition from Hansen et. al. that the model used is not ideal for the system of interest - as well as other work done at Modern Electron (not reported here) which showed little agreement between the plasmatron model and results from PIC simulations - this article will be focused on the broader assumptions made about the plasmatron and specifically how plasma resistance impacts its performance. The biggest of these assumptions is that the electron and ion temperatures are uniform in the interior of the plasmatron device and equal to the cathode temperature. Seeing as the electron and ion temperatures are assumed to be equal and both species are assumed to be Maxwellian, there should be no impact from Coulomb collisions in this model. However, plasma resistance is added in this model as a shift in the device output voltage, calculated as
\begin{equation}
    \Delta V = Jd\eta
\label{eq:plasma_resistance}
\end{equation}
where $J$ is the current density flowing through the plasmatron and $d$ is the inter-electrode gap distance. The plasma resistivity, $\eta$, is calculated from the expression derived in Goldston and Rutherford (1995)\cite{goldstonIntroductionPlasmaPhysics1995}, specifically for the Lorentz-gas (an approximation where ions are assumed to be infinitely massive and only electron-ion collisions are included), namely
\begin{equation}
    \eta = \frac{m_e^{1/2}Ze^2\ln\Lambda}{32\pi^{1/2}\varepsilon_0^2(2k_BT_e)^{3/2}},
\label{eq:plasma_resistivity}
\end{equation}
where $Z$ is the ionization level of ions in the system, $e$  the charge of an electron, $m_e$ its mass, $\ln\Lambda$ is the Coulomb logarithm, $\varepsilon_0$ the permittivity of free space, $k_B$ Boltzmann's constant and $T_e$ the electron temperature. Goldston and Rutherford noted that based on the work by Spitzer and Harm (1953)\cite{PhysRev.89.977} the resistivity should be increased by a factor of $1.7$ to account for electron-electron scattering. This correction is included in the Hansen et. al. plasmatron model. According to the NRL Plasma Formulary\cite{NRL_formulary}, the Coulomb logarithm, $\ln \Lambda$, can be approximated with the following equation for systems where $T_im_e/m_i < T_e < 10 Z^2$ (which is the case in the plasmatron)
\begin{equation}
    \ln\Lambda = 23 - \ln\left(Z\sqrt{\frac{n_e}{T_e^3}}\right).
\label{eq:coulomb_log}
\end{equation}
In the Coulomb logarithm expression above, temperatures are in eV and the electron density, $n_e$, in cm$^{-3}$.\\

The central question this article is aimed at answering is whether the approach of Hansen et. al. to add a voltage shift to the plasmatron JV-curve (calculated via Eq.~\ref{eq:plasma_resistance}) is appropriate to account for Coulomb scattering in such devices.

\subsection{Methods}

A modified version of the PIC code \textit{Warp}\cite{warp,friedmanThreeDimensionalParticle1992,groteWARPCodeModeling2005} was used to perform the simulations discussed in this article. The modifications to the code have mostly already been described in Ref.~\cite{PhysRevE.103.023207}. It includes addition of a Monte Carlo collision (MCC)\cite{birdsallParticleincellChargedparticleSimulations1991} module to capture collision events between electrons or ions and neutral background particles (see the supplemental material of Ref.~\cite{PhysRevE.103.023207} for a benchmark of the implementation versus the results of Turner et. al. (2013)\cite{turnerSimulationBenchmarksLowpressure2013}). Scattering from argon background gas of 5 Torr at 725 $^o$C was included in all simulations discussed here, including elastic, excitation and ionization scattering events for electrons and elastic-, back-scattering and charge exchange for ions. The cross-sections for these collision events were obtained from Phelps (1994)\cite{phelpsApplicationScatteringCross1994} through the LXCat database. A specialized electrostatic solver was also added to \textit{Warp} that directly solves Poisson's equation for the geometry used here, by leveraging superLU\cite{li05} to decompose the finite difference matrix which describes the linear system. Finally, a Coulomb collision module to handle electrons scattering off ions was added to the code and is described in detail below. All simulations were done with a constant electron injection rate from the left domain edge of the simulation (cathode) of $4.78\times 10^{18}$ electrons/s, which corresponds to $J_{emit} = 8.52$ A/cm$^2$ (the emission current density from a thermionic cathode at 1200 $^o$C with a work-function of 2.1 eV and Richardson constant of $60$ A/cm$^2$/K$^2$ as calculated with the Richardson-Dushman equation\cite{crowellRichardsonConstantThermionic1965}). These thermionically emitted electrons were represented by injection of $24$ appropriately weighted macro-particles per timestep, with positions randomly sampled over the left domain boundary. The injected macro-particles had velocities sampled from the "thermionic emission distribution" derived in the supplemental material of Ref.~\cite{PhysRevE.103.023207}. The mean value of velocity perpendicular to the gap from that distribution is $\bar{v}_z = \sqrt{\pi k_BT/(2m)} \approx 1.87\times 10^5$ m/s, giving a local electron density of $J_{emit}/e/\bar{v}_z\approx 2.84\times 10^{18}$ cm$^{-3}$, and corresponding Debye length of roughly $1.57$ $\mu$m (using the cathode temperature). All simulations discussed used a mesh size of $0.75$ $\mu$m to ensure sufficient resolution of that length scale. A timestep of $9.58\times 10^{-13}$~s was used in all simulations to ensure that the CFL-condition would be met for electrons with up to $3$ eV of kinetic energy. This number was chosen to be roughly $10$ times larger than the typical energies encountered in the system. Simulations were performed with different numbers of cells in the $z$-direction, corresponding to different inter-electrode gap distances including 0.5 mm, 0.75 mm and 1 mm. All simulations were done in 2d with 12 cells in the $x$-direction. This was done (as opposed to 1d simulations) due to the findings from Ref.~\cite{PhysRevE.103.023207} that the plasma dynamics in thermionic converters are affected by the dimensionality of the simulation. Dirichlet boundary conditions were used in the $z$-direction and periodic boundary conditions in the $x$-direction.
In accordance with the plasmatron model, from Hansen et. al., electron-ion pairs were continuously injected into the simulation at random positions (uniformly sampled over the simulation domain). The velocity of injected particles were sampled from a Maxwell distribution with temperature equal to the cathode temperature (1200 $^o$C) for the electrons and 725 $^o$C for the ions. For each gap setting, two sets of simulations were performed with different volumetric injection rates ($9.82\times 10^{15}$ and $3.27\times 10^{16}$ ions/electrons per second) in order to achieve different steady state plasma densities. For computational performance reasons, this injection of particles occurred every 5 simulation steps by injection of 20 appropriately weighted ion macro-particles and 20 electron macro-particles. In order to reach steady state in a reasonable number of computational steps, the simulations were all pre-seeded with a neutral plasma with the estimated steady state density that would result from the given injection rate, $S$. This expected steady state density, $\tilde{n}$, was estimated according to $$\tilde{n}=SA\sqrt{\frac{2\pi m_i}{k_B T_i}},$$ where $A$ is the area of the simulation domain boundary (0.09 cm$^2$), $m_i$ the ion mass, $T_i$ the ion temperature (725 $^o$C) and $k_B$ Boltzmann's constant. The initial seed plasma consisted of 200 macro-particles per cell of both ion and electron species types. The simulations were run up to $15$ $\mu$s in simulation time ($\approx$ 16 million steps) in order to capture multiple crossing times for the ions. It was confirmed that the simulations reached a state where the ion loss rate equaled the ion injection rate and the plasma density stabilized (see the supplemental material). The total electronic charge collected on the anode over the final several thousand simulations steps were tracked to provide the steady state diode current.

\subsubsection{Coulomb Scattering}

The PIC approach presented by Manheimer et. al. (1997)\cite{MANHEIMER1997563} was followed to capture Coulomb collisions, but in order to keep the computational load to a minimum, only electron-ion scattering was included. This is justified by the fact that all ions in the simulation are injected with velocities sampled from a Maxwellian distribution, hence ion-ion collisions would have no impact on the system dynamics. Ion-electron collisions are neglected on the bases that electron to ion momentum transfer is much slower (by a factor of $m_i/m_e$) than the reverse process\cite{goldstonIntroductionPlasmaPhysics1995}, and seeing as the expected ion lifetime in the simulation is only a factor of $\sqrt{m_i/m_e}$ longer than the electron lifetime, the expectation is that ions would be lost to the electrodes before undergoing significant momentum transfer with electrons. The electron-electron relaxation period is similar to that of electron-ion collisions, but seeing as the temperature difference between the electrons and ions are much greater than between the Maxwellian electrons in the system and the thermionically emitted ones, it is still expected that the dominant effect will be due to electron-ion collisions. It would certainly be worthwhile to add electron-electron collisions in a follow-up work.

Focusing on Coulomb scattering of electrons off ions, the Fokker-Planck equation\cite{krall1973principles,MANHEIMER1997563} for this process is used as the starting point.
\begin{equation}
    \left(\frac{\partial f_e}{\partial t}\right)_{coll} = -\frac{\partial}{\partial\mathbf{v}}\cdot\mathbf{F}_d(\mathbf{v})f_e(\mathbf{v})+\frac{1}{2}\frac{\partial^2}{\partial\mathbf{v}\partial\mathbf{v}}:\mathbf{D}(\mathbf{v})f_e(\mathbf{v})
\label{eq:fokker_planck}
\end{equation}
where $f_e$ is the electron velocity distribution (for clarity the $:$ represents the double dot product). The dynamical friction coefficient, $\mathbf{F}_d$, and diffusion tensor, $\mathbf{D}$, are given by,
\begin{subequations}
  \begin{align}
    \mathbf{F}_d(\mathbf{v}) &= \frac{nZ^2e^4}{2\pi\varepsilon_0m_e^2}\ln\Lambda\frac{\partial H}{\partial\mathbf{v}}\\
    \mathbf{D}(\mathbf{v}) &= \frac{nZ^2e^4}{2\pi\varepsilon_0m_e^2}\ln\Lambda\frac{\partial^2 G}{\partial\mathbf{v}\partial\mathbf{v}}
  \end{align}
\label{eq:dyn_coefficients}
\end{subequations}
where $n$ is the density of ions, $Z$ is the charge state of the ions, $e$ is the electronic charge, $m_e$ the electron mass and $\varepsilon_0$ the permittivity of free space. Evaluating these quantities requires solving the so-called Rosenbluth potentials, given by
\begin{subequations}
  \begin{align}
    H(v) &= \left(1+\frac{m_e}{m_i}\right)\int d^3\mathbf{\tilde{v}}\frac{f_i(\mathbf{\tilde{v}})}{|\mathbf{v}-\mathbf{\tilde{v}}|}\\
    G(\mathbf{v}) &= \int d^3\mathbf{\tilde{v}}f_i(\mathbf{\tilde{v}})|\mathbf{v}-\mathbf{\tilde{v}}|
  \end{align}
\label{eq:full_rosenbluth}
\end{subequations}
where $f_i(\mathbf{\tilde{v}})$ is the ion velocity distribution function and $m_i$ the ion mass. Following the approach from Manheimer et. al. (1997)\cite{MANHEIMER1997563} these integrals can be simplified under the assumption that the ions are infinitely heavy compared to the mass of an electron. In this case, no energy is transferred between electrons and ions during a scattering event but the electron velocity is simply rotated. Furthermore, the ion distribution function (for the sake of solving the Rosenbluth potentials) can be written as $f_i(\mathbf{v})=\delta(\mathbf{v})$, which allows a closed form solution of the integrals in Eq.~\ref{eq:full_rosenbluth}, which can then be used to evaluate the quantities from Eq.~\ref{eq:dyn_coefficients}, giving
\begin{subequations}
  \begin{align}
    \mathbf{F}_d(\mathbf{v}) &= \frac{nZ^2e^4}{2\pi\varepsilon_0m_e^2v^2}\ln\Lambda\begin{bmatrix}
0\\
0\\
1
\end{bmatrix} \\
    \mathbf{D}(\mathbf{v}) &= \frac{nZ^2e^4}{2\pi\varepsilon_0m_e^2v}\ln\Lambda \begin{bmatrix}
1 & 0 & 0\\
0 & 1 & 0\\
0 & 0 & 0
\end{bmatrix}
  \end{align}
\label{eq:final_coefficients}
\end{subequations}
in a coordinate system that has been rotated such that $\mathbf{v}=v\begin{bmatrix}0 & 0 & 1\end{bmatrix}^T$.
In order to implement this in a PIC code, Manheimer et. al. (1997)\cite{MANHEIMER1997563} uses a Langevin equation equivalent to Eq.~\ref{eq:fokker_planck}\cite{Tabar2019}, namely
\begin{equation}
    \Delta\mathbf{v}=\mathbf{F}_d\Delta t + \mathbf{Q}
\label{eq:langevin}
\end{equation}
where $\mathbf{Q}$ is a random vector with components sampled from normal distributions:
\begin{equation}
    \mathbf{Q} = \sqrt{\Delta t}
\begin{bmatrix}
\mathcal{N}(0, \sqrt{\mathbf{D}_{11}}) \\ \mathcal{N}(0, \sqrt{\mathbf{D}_{22}}) \\ \mathcal{N}(0, \sqrt{\mathbf{D}_{33}})
\end{bmatrix}.
\label{eq:dv_distribution}
\end{equation}
Seeing as the scattering of ions are neglected in this treatment of Coulomb collisions, in order to force energy conservation, only the two transverse components of $\mathbf{Q}$ were sampled while the third component was calculated to preserve the magnitude of the velocity vector for each electron during a scattering event. In this line the dynamical friction coefficient was also ignored. As already noted, with this choice of implementation there is no thermalization due to Coulomb collisions, but instead it only serves to isotropize the electron velocity distribution function. The actual implementation of this process in \textit{Warp} was as follows: An extra simulation step was inserted at the end of the PIC cycle where the velocity of each electron was perturbed based on a sampled $\mathbf{Q}$ (with third component calculated as described to conserve energy). See supplemental material for details of how the velocity perturbation was done. The calculation of $\mathbf{D}$ requires knowledge of extensive (grid) quantities such as the electron and ion density and electron temperature (to calculate the Coulomb logarithm from Eq.~\ref{eq:coulomb_log}). These values were calculated on the grid and interpolated to the electron positions before calculating $\mathbf{D}$ for each electron. In order to save computational effort, the values of $\mathbf{D}$ were only updated once every 5 simulation steps. This subcycling proved to give valuable savings in computational time without affecting the simulation results (see supplemental material). A benchmark to test the accuracy of the implementation against an analytic solution of the Fokker-Planck equation has been included in the supplemental material as well.

\subsection{Results}

\begin{figure}[b]
    \centering
    \includegraphics[width=1.0\columnwidth]{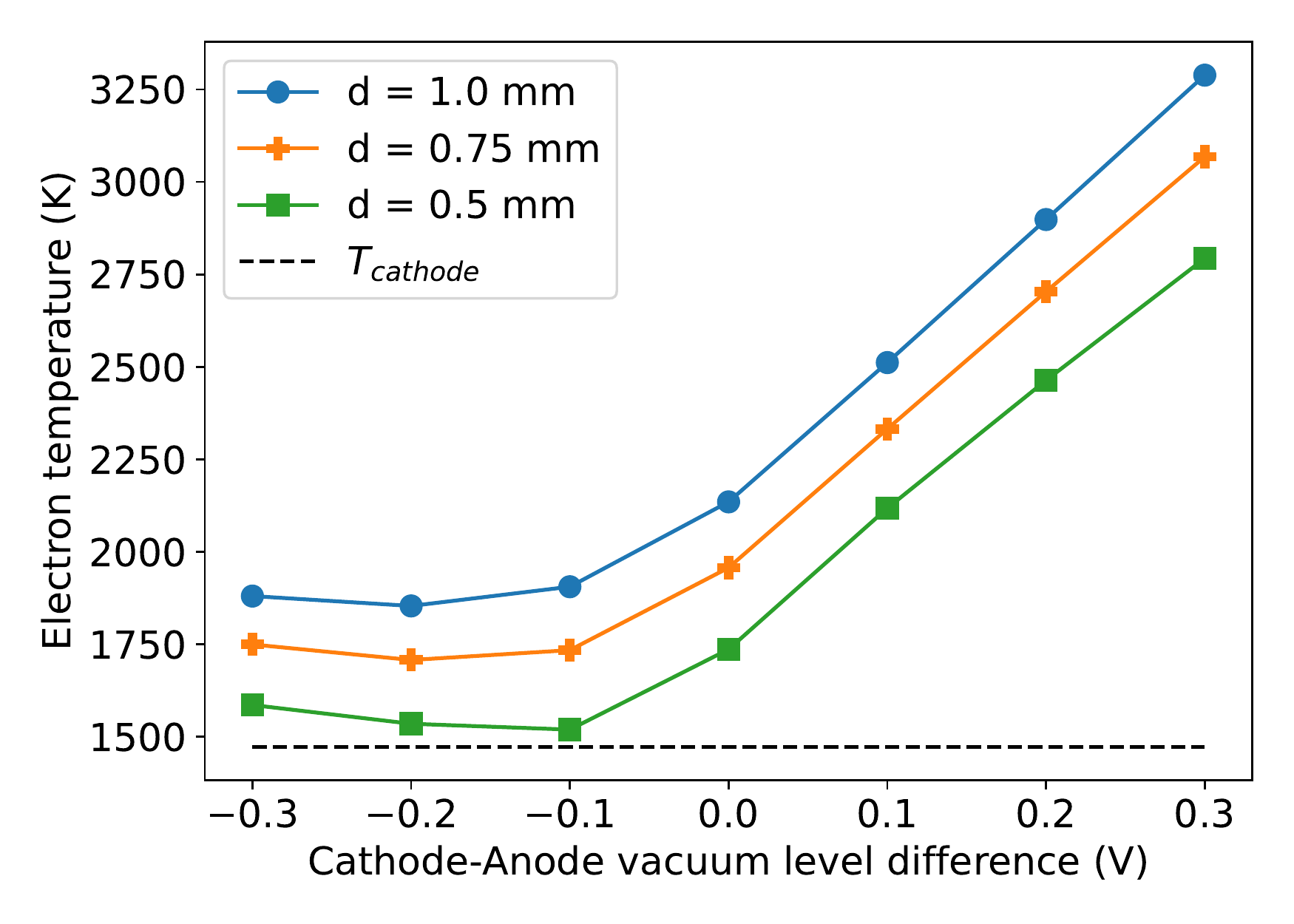}
    \caption{Average electron temperature for different inter-electrode gap distances with fixed volumetric injection rate of $3.27\times 10^{16}$ particles/s and no Coulomb scattering. The simulated cathode temperature is also shown for comparison.}
    \label{fig:elec_temps}
\end{figure}

Firstly, the assumption made by Hansen et. al. (1976)\cite{plasmatron} that the electron temperature equals the cathode temperature can be assessed based on the PIC results. The average electron temperature for simulations done at different gaps and relative electrode potentials are shown in Fig.~\ref{fig:elec_temps}. The electron temperature was calculated as $T_e = \frac{2}{3}\langle K_E\rangle/k_B$, where $\langle K_E\rangle$ is the average kinetic energy for all electrons in the simulation. The PIC simulations show a dependence of the electron temperature on the inter-electrode gap and electrode biases (both neglected in the plasmatron model). With the $T^{-3/2}$ factor in calculating the plasma resistivity, a factor of 2 difference in electron temperature would result in an approximately $65\%$ decrease in plasma resistivity, already indicating that the results from Hansen et. al. likely overestimated the impact of plasma resistance on the plasmatron performance.

\begin{figure}
    \centering
    \includegraphics[width=1.0\columnwidth]{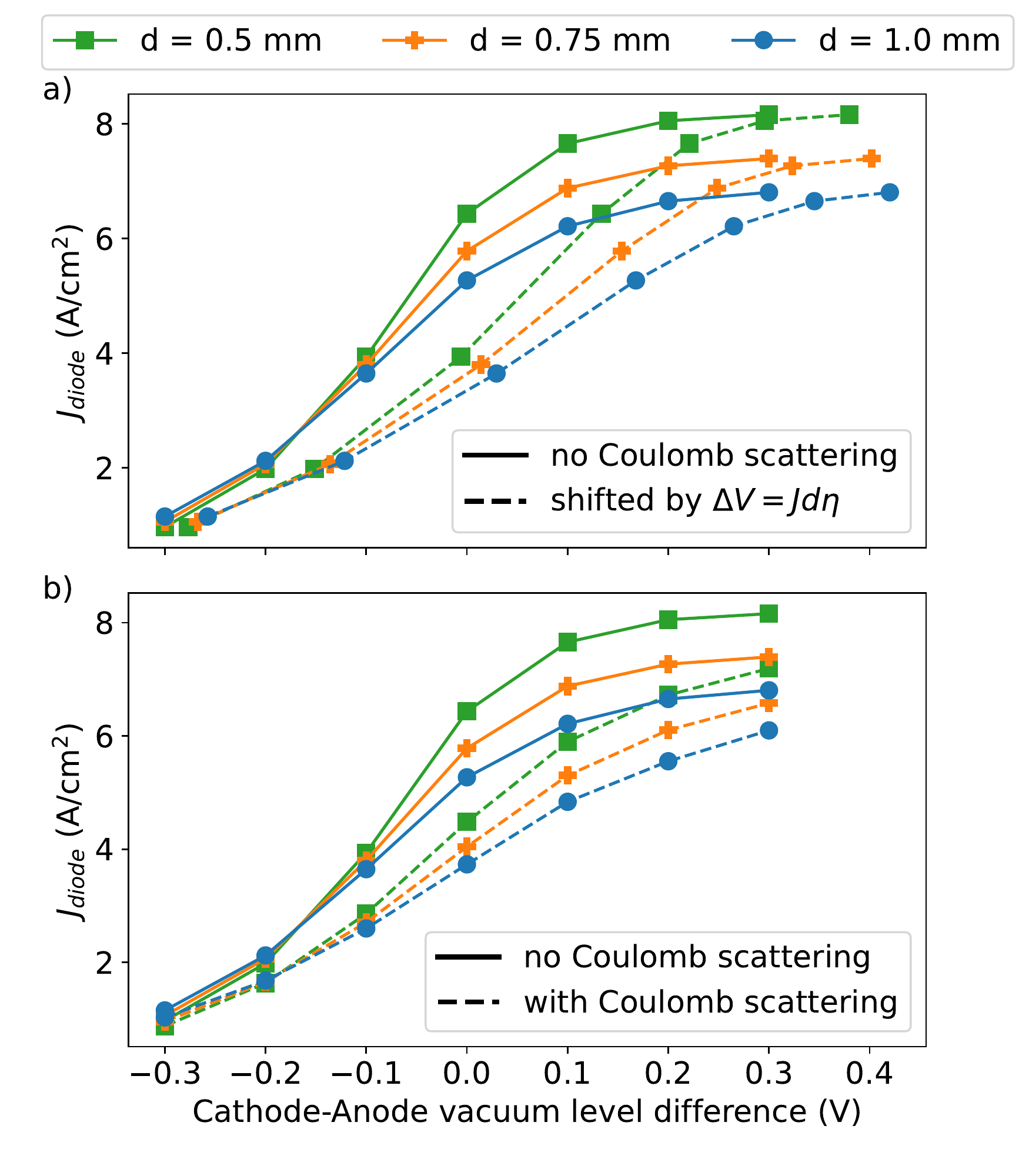}
    \caption{Simulated JV-curves for different inter-electrode gap distances with fixed volumetric injection rate of $3.27\times 10^{16}$ particles/s. The adjusted IV-curves after shifting the voltage (see Eq.~\ref{eq:plasma_resistance}) based on the calculated plasma resistivity for each point (from Eq.~\ref{eq:plasma_resistivity}) is also shown (a) as well as the impact on the JV-curves when including Coulomb scattering in the simulation (b).}
    \label{fig:iv_curves}
\end{figure}

\begin{figure}
    \centering
    \includegraphics[width=1.0\columnwidth]{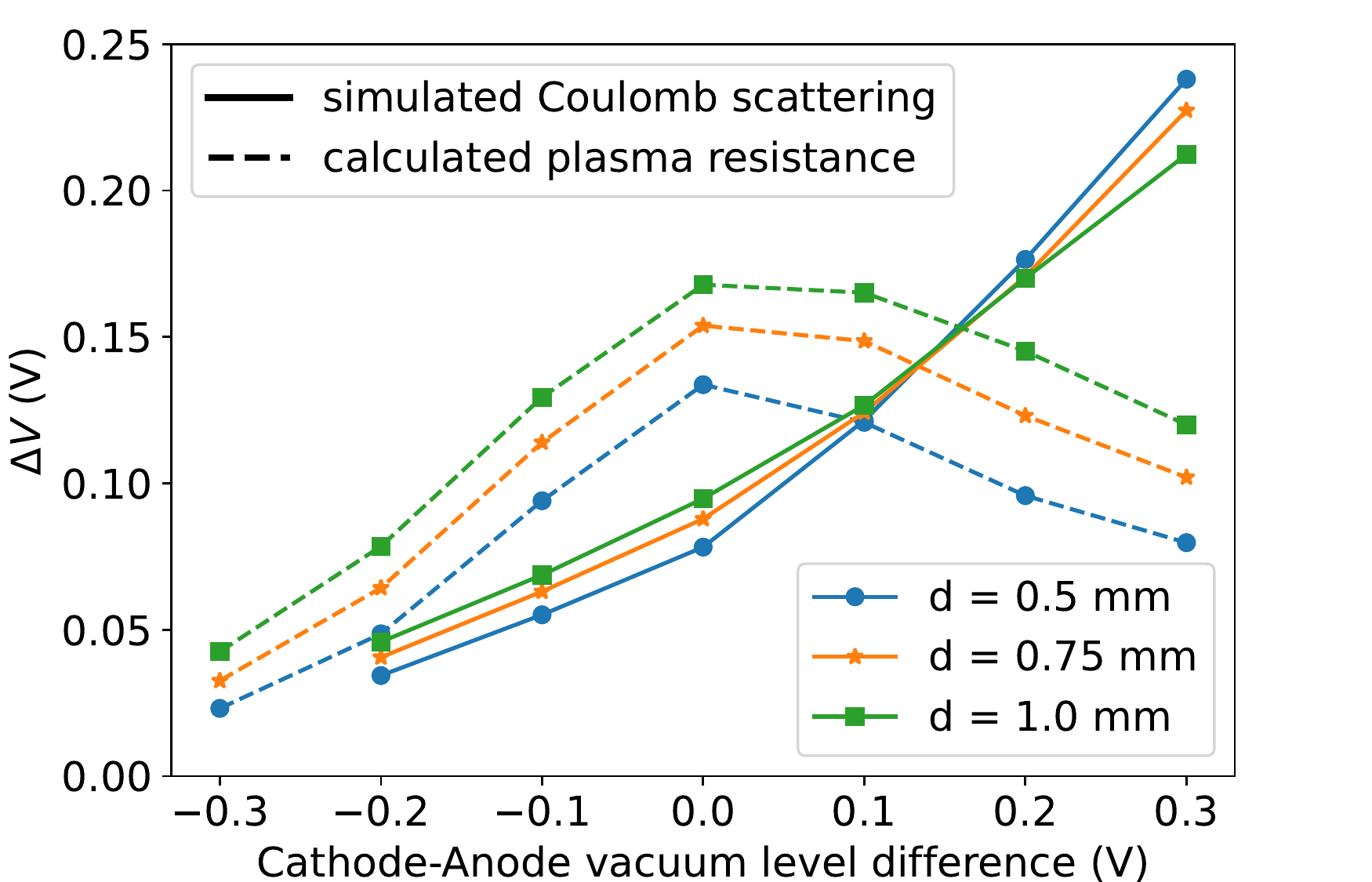}
    \caption{Comparison of the voltage drop incurred in the thermionic converter due to plasma resistance as calculated with Eq.~\ref{eq:plasma_resistance} (dashed) versus PIC simulations with Coulomb scattering included (solid) for different gaps with fixed volumetric injection rate of $3.27\times 10^{16}$ particles/s.}
    \label{fig:V_shift_compared}
\end{figure}

Seeing as the electron temperature was found to vary substantially with system parameters, all further analysis will deviate from the assumption that $T_e = T_{cathode}$ and instead use the resulting electron temperature from the PIC simulations when calculating plasma resistivity with Eq.~\ref{eq:plasma_resistivity}. The average electron density from the simulation results were also used in those calculations and can be found in the supplemental material. The JV-curves in Fig.~\ref{fig:iv_curves}a show the current density transmitted through the plasmatron without Coulomb scattering included as well as how these curves would shift if the plasma resistance is included in the same approach as taken by Hansen et. al. (1976). It also shows in Fig.~\ref{fig:iv_curves}b the same JV-curves without Coulomb scattering as well as the resulting JV-curves when including Coulomb scattering in the simulation (through the Langevin based method described earlier). Note the discrepancy between the two sets of JV-curves that include plasma resistance.
In order to highlight the difference in impact on the predicted JV-curve of the plasmatron from the two approaches, Fig.~\ref{fig:V_shift_compared} shows the voltage shift incurred from plasma resistance for both approaches. In the case of the simulated Coulomb scattering $\Delta V$ was obtained for a given bias, $V$, by finding the bias, $V^*$ of the JV-curve without Coulomb scattering such that $J_{no Coul}(V^*) = J_{Coul}(V)$, and taking $\Delta V = V-V^*$.
Notice that the increase in $\Delta V$ with increased gap is much less in the case where Coulomb scattering was captured in the simulation than in the case where Eq.~\ref{eq:plasma_resistance} is used. This comparison effectively indicates that the assumption of nearly Maxwellian electrons, made in the derivation of the Lorentz gas resistivity (Eq.~\ref{eq:plasma_resistivity}), is not applicable to the plasmatron system.
All simulations presented above were also performed with a different ion injection rate to see if the results discussed here hold at different plasma densities. Those results are similar to what has been shown here and have been included in the supplemental material for the interested reader.

\subsection{Discussion}

The question of how much the plasma resistance impacts the performance of a plasmatron TEC can now be answered. As discussed in the introduction, the relative vacuum potential levels of the cathode and anode, along with the difference in their work-functions, combine to give the output voltage of the TEC. Therefore, the $JV$-curves presented in Fig.~\ref{fig:iv_curves} can be translated to current density versus output voltage curves ($JV_{out}$-curves) given a difference in electrode work-functions, $\Delta \phi$. The output power of the TEC can also be calculated as a function of relative vacuum levels via
\begin{equation}
    P_{out}(V) = J\cdot V_{out} = J(\Delta\phi-V),
\label{eq:power_out}
\end{equation}
where $V$ is the cathode-anode vacuum level difference (note that from the above it is clear that once $V>\Delta \phi$ the TEC becomes power consuming rather than power producing). To understand the impact plasma resistance has on the performance potential of a TEC, it is necessary to look at the maximum power point (MPP) i.e. the voltage bias at which the TEC produces the most electrical power. This point can be found through simple optimization by finding the value $V^*$ such that $$P'_{out}(V^*)=J'(V^*)[\Delta\phi -V^*]-J(V^*)=0,$$ where the primes represent derivatives with respect to $V$.
\begin{figure}
    \centering
    \includegraphics[width=1.0\columnwidth]{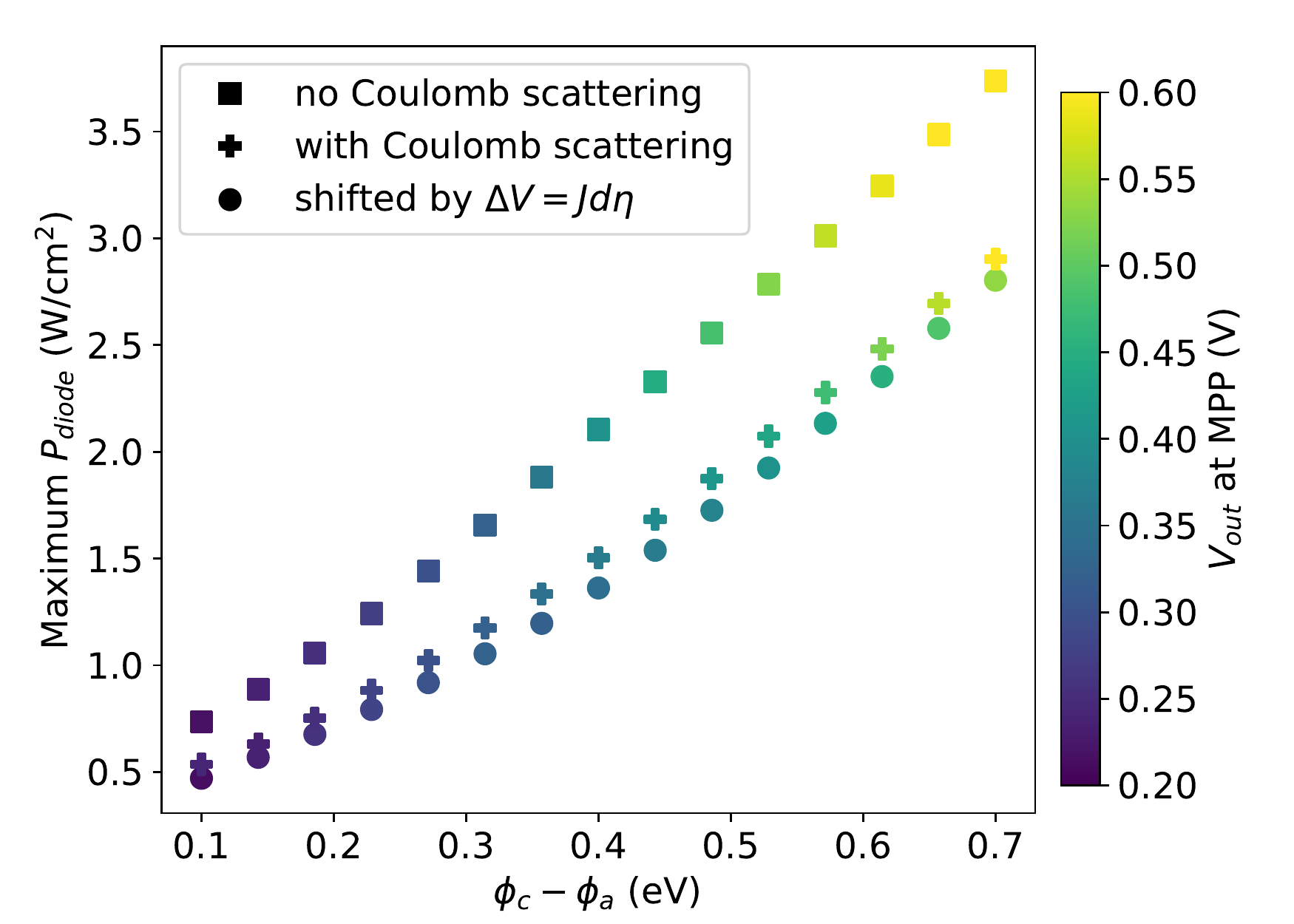}
    \caption{Calculation of the maximum output power as a function of cathode-anode work-function difference for all the JV-curves shown in Fig.~\ref{fig:iv_curves} for the $1$ mm gap case. The output voltage at the maximum power point (MPP) is indicated by the symbol coloring. Note that this calculation neglects the power drain associated with generating the plasma and therefore is not an accurate absolute measure of expected output power density.}
    \label{fig:power_out}
\end{figure}
The result of this optimization is shown in Fig.~\ref{fig:power_out} for the $d = 1$ mm simulations discussed earlier. Interestingly, the two methods of accounting for plasma resistance result in roughly the same maximum output power from the plasmatron but at different output voltages. This is important since the efficiency of a TEC is heavily determined by the device's output voltage\cite{hatsopoulos_thermionic_1973}, with a higher output voltage for the same net power corresponding to a higher efficiency. This can be understood by considering that the heat loss from the cathode due to electron cooling is proportional to the diode current, so a lower current would result in less cooling of the cathode. Therefore, the simulations discussed here indicate that plasmatron devices can operate with higher conversion efficiency than previously thought based on the work from Hansen et. al. (1976).

Another take-away from this work is the following: Hansen et. al. (1976) concluded from their model that the impact of plasma resistance on the plasmatron performance, as expressed with Eq.~\ref{eq:plasma_resistance} and Eq.~\ref{eq:plasma_resistivity} (using $T_e=T_{cathode}$), compels thermionic converter design utilizing an auxiliary plasma generation source to have small gaps - since $\Delta V\propto d$ in their model. They estimated a required output power threshold of $\sim2$ W/cm$^2$ to make TECs attractive for terrestrial applications, which from their calculations require $d<0.5$ mm. The higher fidelity simulation results reported in this article (specifically Fig.~\ref{fig:elec_temps} and Fig.~\ref{fig:V_shift_compared}) show that with increasing gap the electron temperature also increases, which decreases the plasma resistance giving a net $\Delta V$ result that grows much slower than linear with $d$. This means that the plasmatron doesn't necessarily have to be built with a small gap to allow efficient energy conversion.

The question of how electron-electron collisions impact these results is an important one for further investigations of plasmatron-type thermionic converters i.e. ones that use an auxiliary source of high energy electrons to generate a plasma in the inter-electrode gap. Specifically since the collision frequency of such high energy electrons with the relatively low energy thermionically emitted electrons will be much higher than collisions between low energy electrons\cite{goldstonIntroductionPlasmaPhysics1995}. This scattering pathway could influence both the resistivity of the plasma overall and the efficiency with which the high energy electron beam can generate a plasma since rapid thermalization of the high energy beam would negatively impact the rate of collisional ionization events. Such efforts are left to future work.

\section{Acknowledgement}

   The author is grateful to P. Scherpelz for helpful discussions as well as his contributions to the development of the modified version of \textit{Warp}.

%

\renewcommand{\thefigure}{S\arabic{figure}}
\renewcommand{\theequation}{S\arabic{equation}}

\section{Supplemental Material to: Impact of Coulomb Scattering on Argon Plasma Based Thermionic Converter Performance}

\section{Perturbation of electron velocities due to Coulomb scattering}

It is important to keep in mind that the velocity perturbation as specified in the main text assumes a coordinate system in which the particle velocity is $\mathbf{v}=v\begin{bmatrix}0 & 0 & 1\end{bmatrix}$, therefore to apply this perturbation to the particles in the PIC simulation, requires an appropriate rotation. Let the velocity of a simulation particle (in the simulation frame) be $$\mathbf{u} = u_x\mathbf{\hat{x}}+u_y\mathbf{\hat{y}}+u_z\mathbf{\hat{z}},$$ with $$u=|\mathbf{u}|=\sqrt{u_x^2+u_y^2+u_z^2}$$ and define $$u_\perp = \sqrt{u_x+u_y}.$$ Next, let the velocity perturbation (in the rotated frame of $\mathbf{v}$) be $$\mathbf{Q}=\begin{bmatrix}Q_1 & Q_2 & Q_3\end{bmatrix},$$ where $Q_1$ and $Q_2$ have known values after being sampled from the appropriate normal distributions and $$Q_3 = \sqrt{u^2-(Q_1^2+Q_2^2)} - u$$ is fixed to enforce energy conservation. It is necessary to calculate the appropriate rotation to move $\mathbf{Q}$ to the same coordinate system as $\mathbf{u}$, which can be accomplished by calculating the angles, $$\theta = \cos^{-1}\left(\frac{u_z}{u}\right)$$ and $$\phi = \cos^{-1}\left(\frac{u_x}{u_\perp} \right).$$ Finally, the rotation is accomplished via
\begin{equation}
    \mathbf{Q}\mapsto \Delta\mathbf{u} =
\begin{bmatrix}
\cos(\theta)\cos(\phi) & -\sin(\phi) & \sin(\theta)\cos(\phi) \\
\cos(\theta)\sin(\phi) & \cos(\phi) & \sin(\theta)\sin(\phi) \\
-\sin{\theta} & 0 & \cos(\theta)
\end{bmatrix}
\begin{bmatrix}
Q_1 \\ Q_2 \\ Q_3
\end{bmatrix}
\label{eq:rotation}
\end{equation}
and the particle velocity is adjusted according to $$\mathbf{u}' = \mathbf{u} + \Delta\mathbf{u}.$$ In the rare case that $u^2-(Q_1^2+Q_2^2) < 0$ - which can always happen since $Q_1$ and $Q_2$ are sampled from (unbounded) normal distributions - the direction of the particle velocity is uniformly resampled.

\section{Subcycling of Coulomb scattering parameters}

\begin{figure}[h]
    \centering
    \includegraphics[width=1.0\columnwidth]{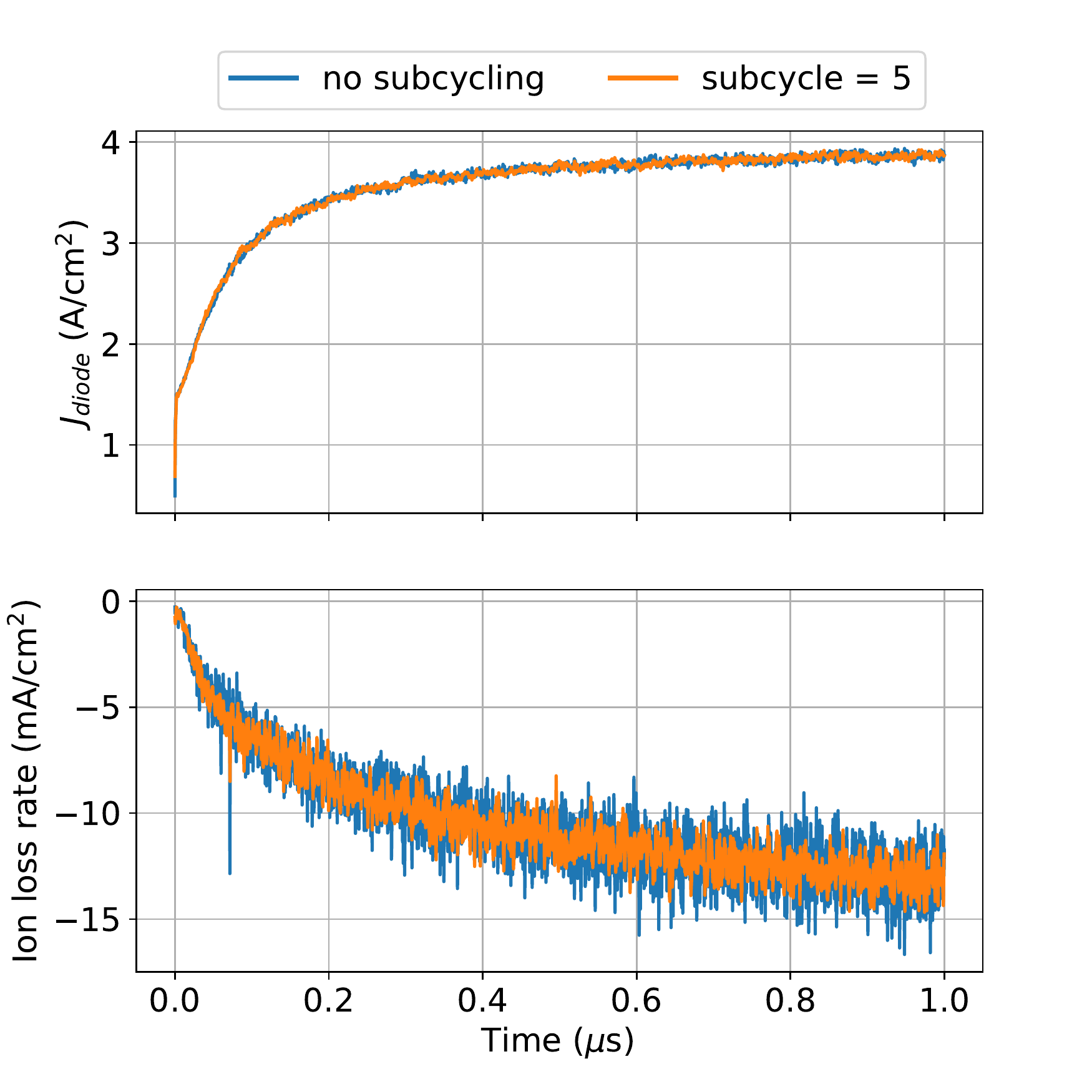}
    \caption{Simulations to assess the impact of subcycling updates to the diffusion tensor. The simulations performed had the cathode-anode vacuum level difference set to 0 V and a particle injection rate of $3.27\times 10^{16}$ particles/s.}
    \label{fig:subcycling}
\end{figure}

As described in the main article, the grid quantities used to calculate the diffusion tensor, $\mathbf{D}$, was updated every 5 simulation steps in order to save on computational load. The impact of this was tested by comparing the results against a simulation in which the grid quantities were updated at every step. Time-traces of the electron and ion currents for both approaches are shown in Fig.~\ref{fig:subcycling}, which shows the subcycling approach does not impact the simulation results.

\section{Benchmark of Langevin based Coulomb scattering implementation}

\begin{figure}
    \centering
    \includegraphics[width=1.0\columnwidth]{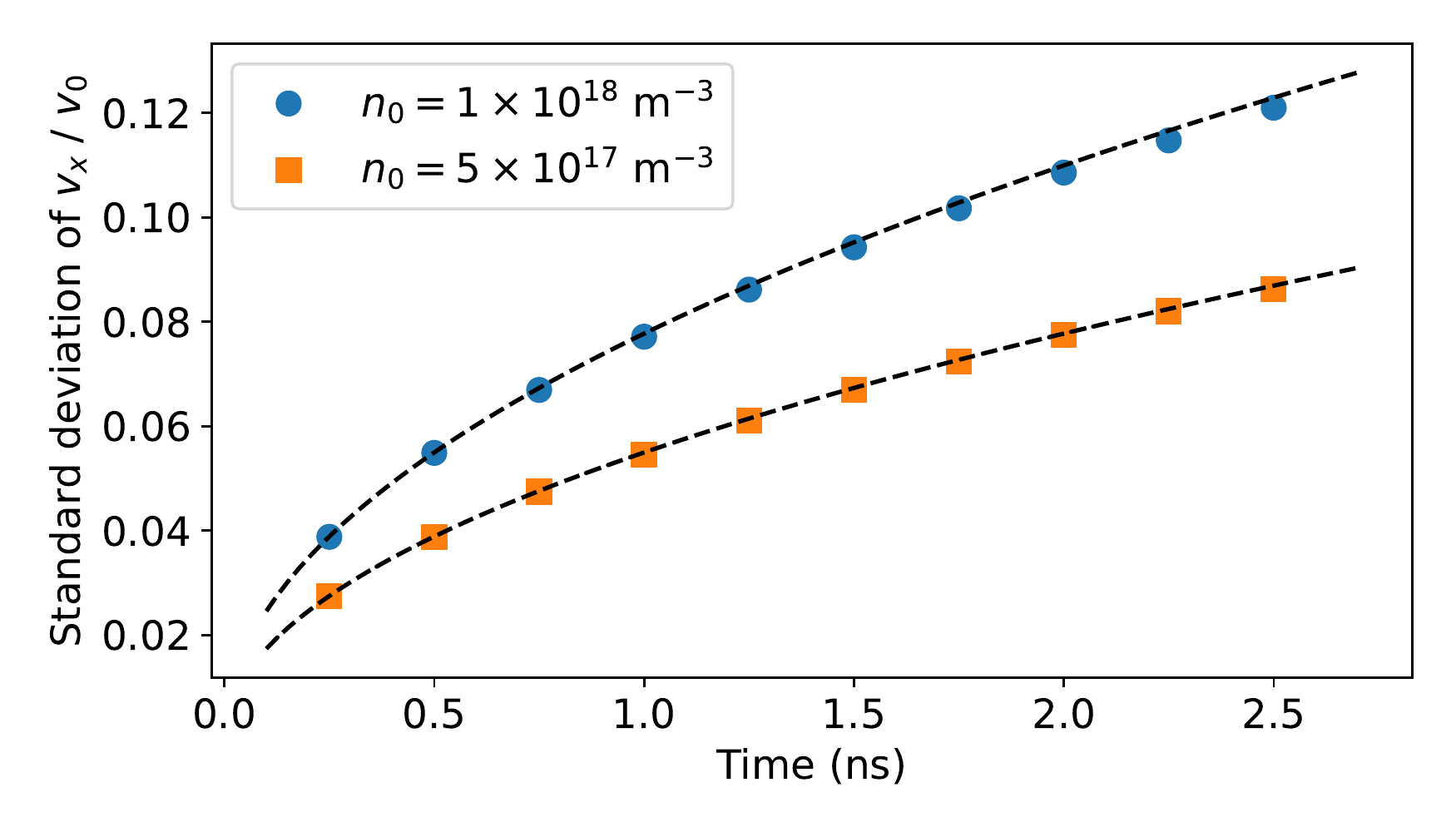}
    \caption{Benchmark calculation of the Langevin Coulomb scattering implementation in \textit{Warp} under the Lorentz gas approximation, as described in the main text. The black dashed line indicates the theoretical solution as given by Eq.~\ref{eq:analytic_sol}.}
    \label{fig:diffusion}
\end{figure}

In order to confirm the electron-ion Coulomb scattering approach, as described in the main text, was implemented correctly, a set of benchmark calculations were performed. The Fokker-Planck equation with the Lorentz gas assumption (as described in the main text) is given by
\begin{equation}
\begin{split}
    \frac{\partial f}{\partial t} &= \frac{1}{2}\frac{\partial^2}{\partial\mathbf{v}\partial\mathbf{v}}:\mathbf{D}(\mathbf{v})f(\mathbf{v})\\
    &= \frac{nZ^2e^4}{4\pi\varepsilon_0m_e^2v}\ln\Lambda \left(\frac{\partial^2}{\partial v_x^2} + \frac{\partial^2}{\partial v_y^2} \right)f_0
\label{eq:fokker_planck_simplified}
\end{split}
\end{equation}
which can be analytically solved for the specific initial distribution function
\begin{equation}
f_0(\mathbf{v})=\delta(v_x)\delta(v_y)\delta(v_0 - v_z),
\label{eq:f_0}
\end{equation}
for short times compared to the isotropization time. The solution is given by,
\begin{equation}
    f(t) = \frac{1}{2\pi Dt}\exp\left[-\frac{v_x^2}{2Dt}-\frac{v_y^2}{2Dt} \right]\delta(v_0 - v_z)
\label{eq:analytic_sol}
\end{equation}
where $$D = \frac{n_0Z^2e^4}{4\pi\varepsilon_0m_e^2v_0}\ln\Lambda.$$
Note that this solution is only accurate for times, $t$, such that the $v_z$ component of the velocity is still the dominant component. A numerical comparison to this analytic solution of the distribution function is shown in Fig.~\ref{fig:diffusion}, where simulations were performed with different plasma densities and the velocity isotropization rate compared to the prediction.
The benchmark simulations used 1~$\mu$m resolution with a grid of 500x8 cells, periodic boundary conditions in all directions and $\Delta t = 7\times10^{-14}$ s. A fixed Coulomb logarithm with value 7.5 was used. The simulations were seeded with 40000 macro-particles each of electrons and ions randomly placed. The ions were given a mass of 1 kg, to match the Lorentz gas approximation. The electrons had an initial velocity distribution function given by Eq.~\ref{eq:f_0} with $v_0 = 10^6$ m/s.

\section{Simulations reached steady state}

\begin{figure}
    \centering
    \includegraphics[width=1.0\columnwidth]{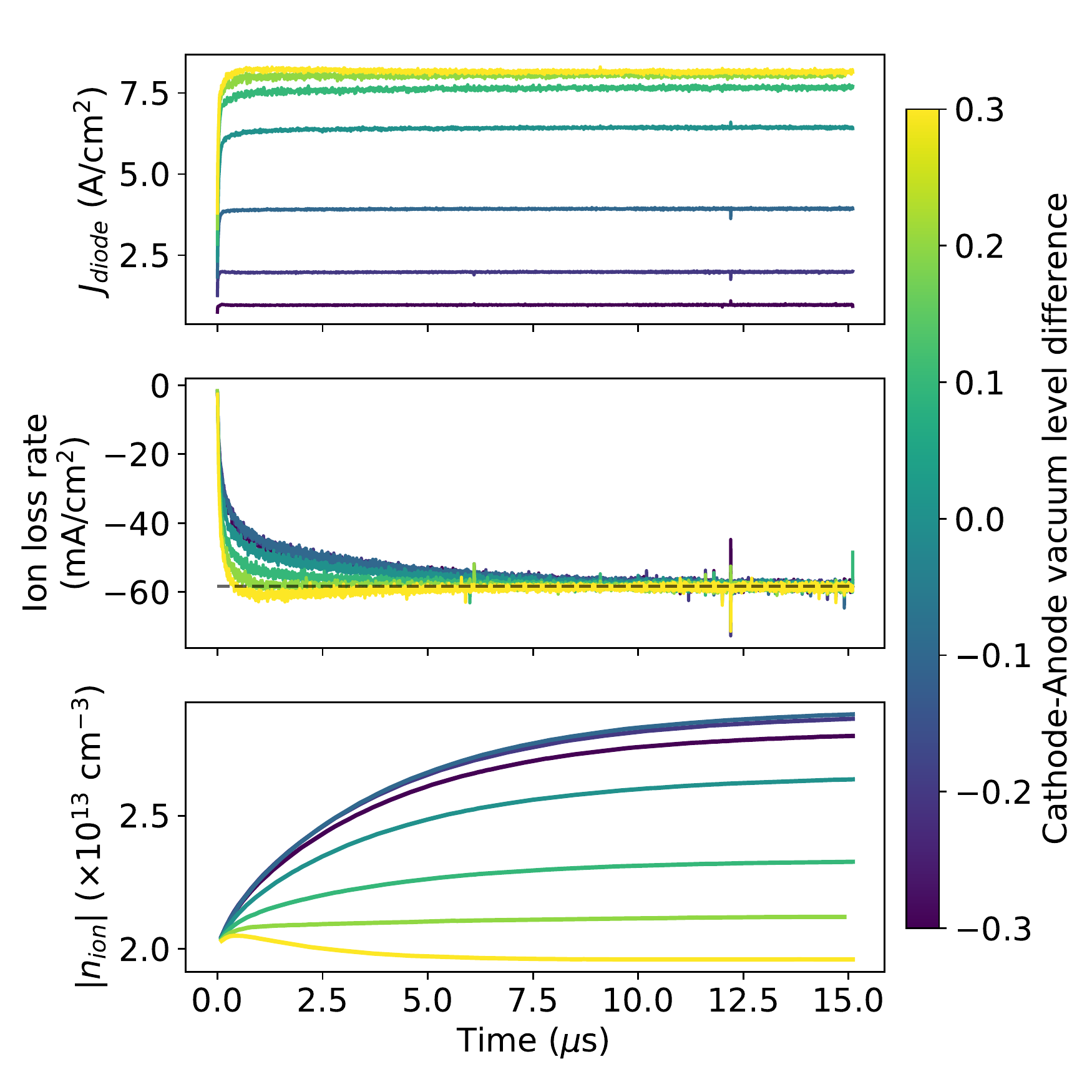}
    \caption{Time traces of the electron flux through the diode (top), ion flux collected on both the cathode and anode (middle) and average ion density (bottom), for a representative simulation with $d = 0.5$ mm and volumetric particle injection rate of $3.27\times 10^{16}$ particles/s. The dashed line in the middle frame shows the injection rate of ions normalized to the cathode area.}
    \label{fig:steady_state}
\end{figure}

Fig.~\ref{fig:steady_state} shows a representative case of the time-evolution of fluxes observed in the simulations as well as the average ion density. The figure shows that the simulation reached a steady state wherein the ion flux out of the system equals the injection rate and the ion density no-longer changed over time. This steady state performance of the system is what was studied in the main manuscript.

\section{Electron density}

\begin{figure}[h]
    \centering
    \includegraphics[width=1.0\columnwidth]{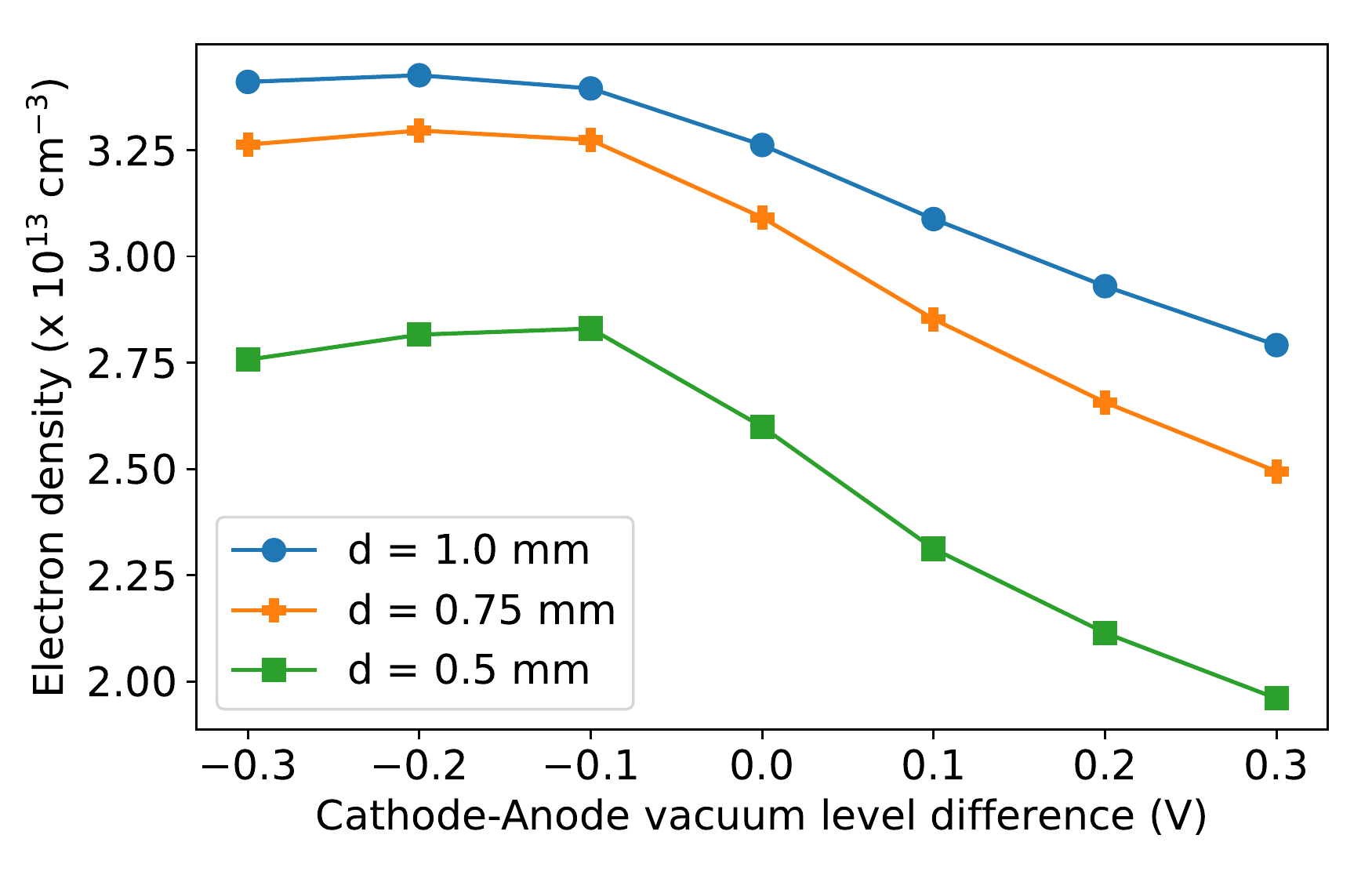}
    \caption{Electron density at different cathode-anode vacuum level differences for the simulations with particle injection rate of $3.27\times 10^{16}$ particles/s.}
    \label{fig:elec_dens}
\end{figure}
The electron density used in the main text when calculating plasma resistivity and the Coulomb logarithm is shown in Fig.~\ref{fig:elec_dens} below for reference. It is expected that the steady-state plasma density would vary based on the relative vacuum level potentials of the cathode and anode since that affects the plasma sheaths in front of the electrodes which determines the ion loss rate from the bulk of the plasma. The simulation setup used here is (with the ion injection rate independent of the electrode biases) has been deemed appropriate since it represent a system where the plasma generation is independent of the state of the thermionic diode.

\section{Results from lower injection rate}

\begin{figure}[h]
    \centering
    \includegraphics[width=1.0\columnwidth]{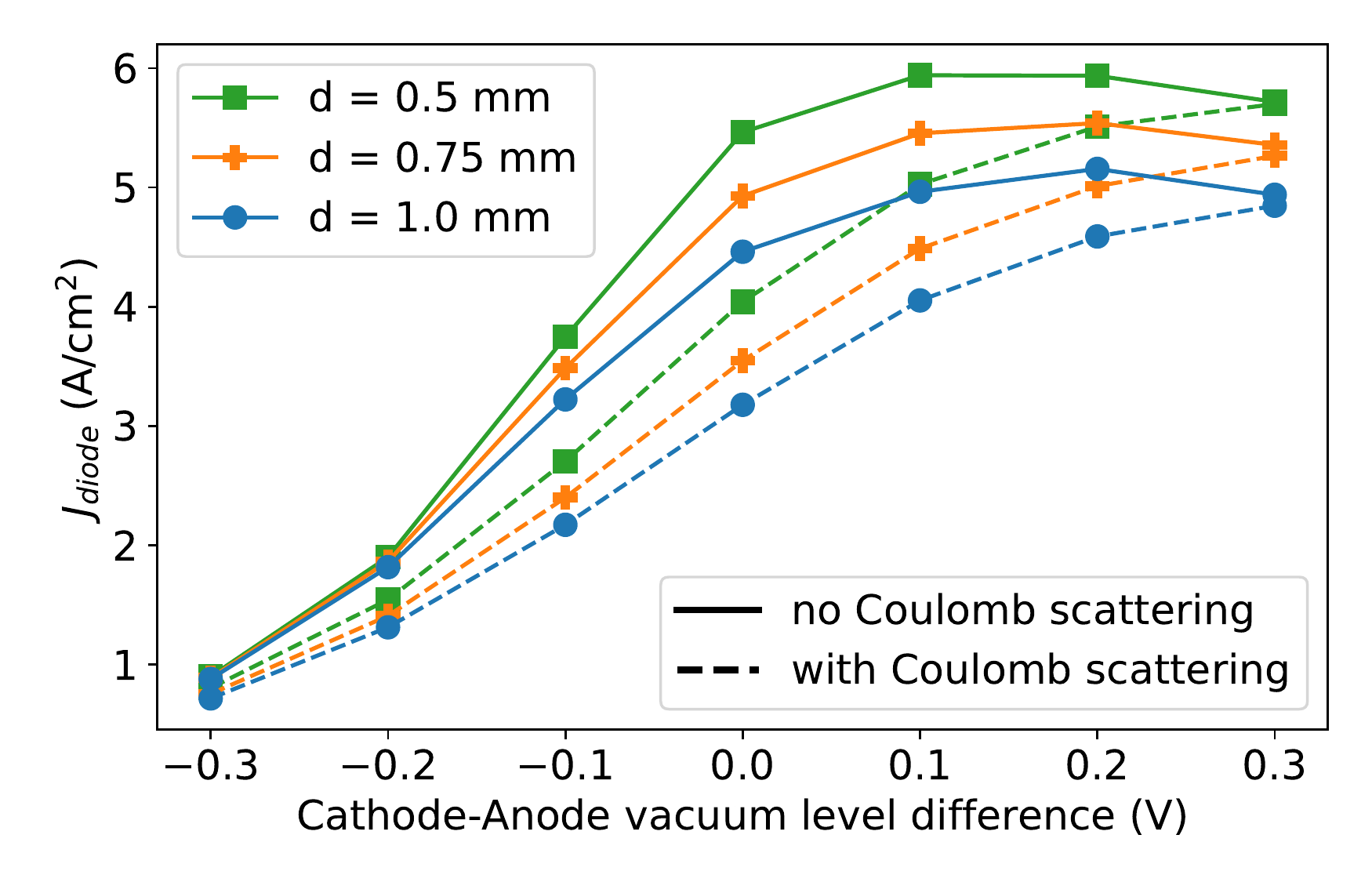}
    \caption{Simulated JV-curves for different inter-electrode gap distances with fixed volumetric injection rate of $9.82\times 10^{15}$ particles/s. The impact on the JV-curves when including Coulomb scattering in the simulation is shown.}
    \label{fig:low_n_ivs}
\end{figure}
As a minimal check that the results discussed in the main article are not specific to the exact system parameters used, the simulations were repeated at a lower injection rate of $9.82\times10^{15}$ particles/s. The results of these simulations are presented here for reference but the conclusions from the main article hold for this data set as well. The JV-curves with and without Coulomb scattering included are shown in Fig.~\ref{fig:low_n_ivs}.
\begin{figure}
    \centering
    \includegraphics[width=1.0\columnwidth]{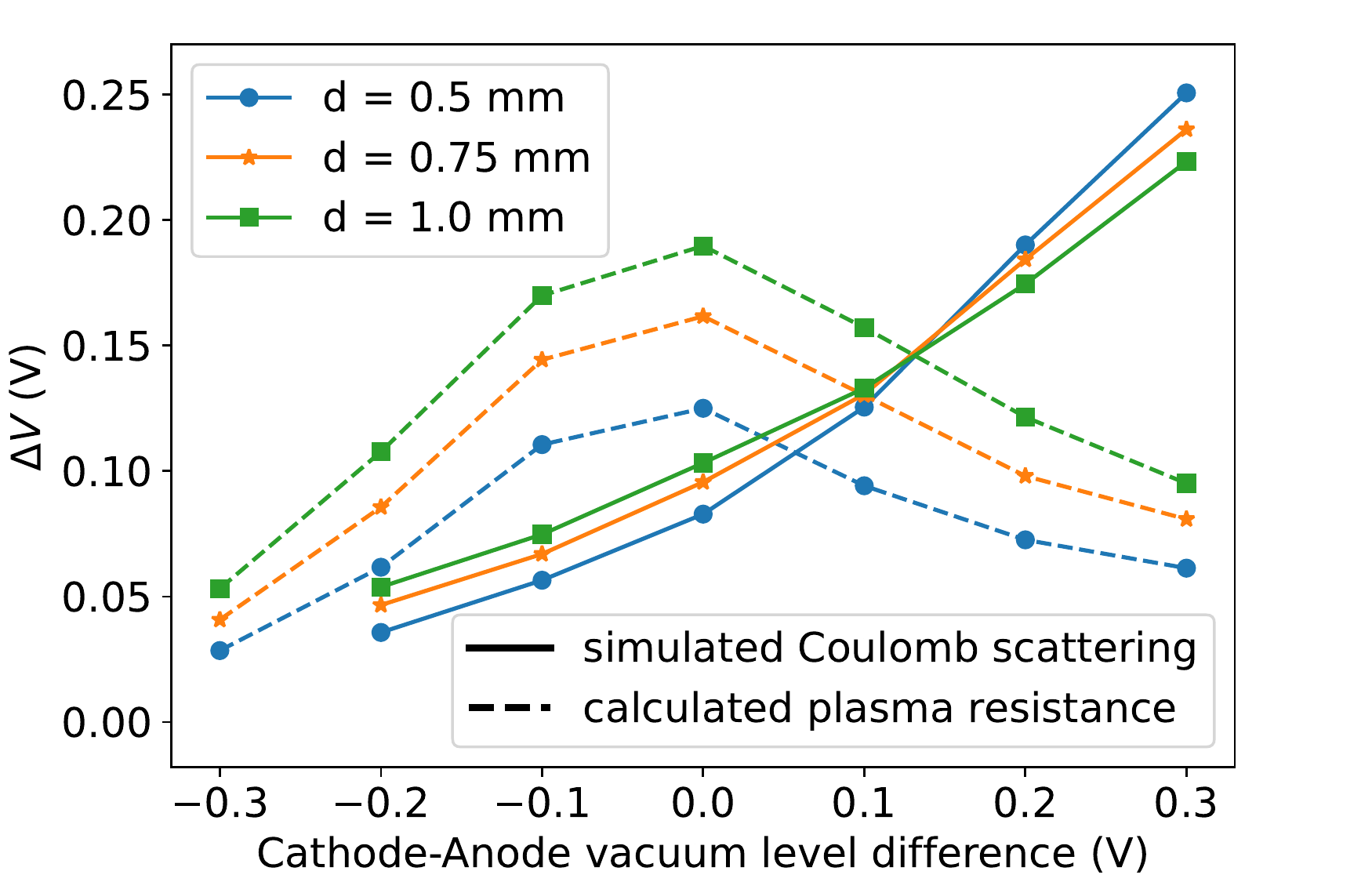}
    \caption{Comparison of the voltage drop in the thermionic converter due to plasma resistance as calculated with Eq.~1 from the main text (dashed) versus PIC simulations with Coulomb scattering included (solid) for different gaps with fixed volumetric injection rate of $9.82\times 10^{15}$ particles/s.}
    \label{fig:low_n_dV}
\end{figure}
A comparison of the shift in JV-curve due to Coulomb scattering (included in the simulation via the Langevin approach discussed in the main text) versus a calculated shift using the Lorentz gas resistivity (main text Eq.~1) is shown in Fig.~\ref{fig:low_n_dV}. Again the higher fidelity approach shows a slower increase in $\Delta V$ with increasing gap than the simplified approach.
Finally, Fig.~\ref{fig:low_n_power} shows the maximum output power for the JV-curves from Fig.~\ref{fig:low_n_ivs} for the $1$ mm gap simulations as a function of cathode-anode work-function difference, similar to Fig. 5 from the main text. The maximum power curve is also shown as calculated with the simplified plasma resistivity treatment discussed in the main text.
\begin{figure}
    \centering
    \includegraphics[width=1.0\columnwidth]{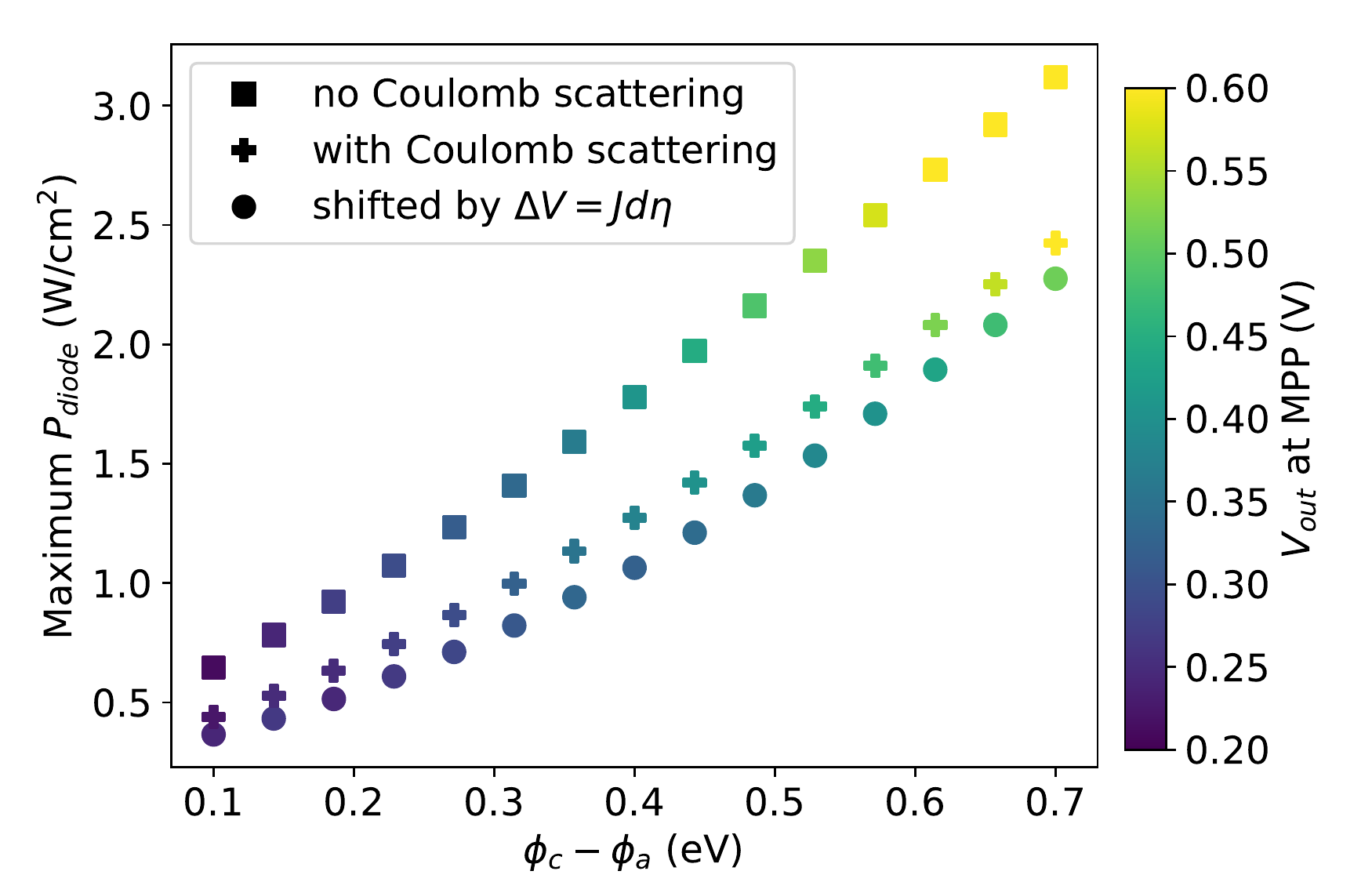}
    \caption{Calculation of the maximum output power as a function of cathode-anode work-function difference for the JV-curves shown in Fig.~\ref{fig:low_n_ivs} for the $1$ mm gap case. The output voltage at the maximum power point (MPP) is indicated by the symbol coloring. Note that this calculation neglects the power cost of generating the plasma and therefore is not an accurate absolute measure of output power density.}
    \label{fig:low_n_power}
\end{figure}
Similarly as with the simulations reported there, the two methods of accounting for plasma resistance do not differ much in the value of maximum output power density (the higher fidelity method does show a slightly higher optimal), but they do differ in the output voltage at which the maximum power is generated. As already discussed, this affects the energy conversion efficiency of the device.

\bibliography{bibliography}

\end{document}